\documentclass[aps,showpacs,preprintnumbers,amsmath,amssymb]{revtex4}
\usepackage{dcolumn}
\usepackage[dvips]{epsfig}
\usepackage{graphicx}
\usepackage{amssymb}
\usepackage{color}
\usepackage{enumerate}

\begin{document}
\baselineskip=0.8 cm
\title{ Chaos in the motion of a test scalar particle coupling to Einstein tensor in Schwarzschild-Melvin black hole spacetime}
\author{Mingzhi Wang$^{1,2}$,  Songbai Chen$^{1,2,3}$\footnote{Corresponding author: csb3752@hunnu.edu.cn}, Jiliang
Jing$^{1,2,3}$ \footnote{jljing@hunnu.edu.cn}}
\affiliation{$ ^1$Institute of Physics and Department of Physics, Hunan
Normal University,  Changsha, Hunan 410081, People's Republic of
China \\ $ ^2$Key Laboratory of Low Dimensional Quantum Structures \\
and Quantum Control of Ministry of Education, Hunan Normal
University, Changsha, Hunan 410081, People's Republic of China\\
$ ^3$Synergetic Innovation Center for Quantum Effects and Applications,
Hunan Normal University, Changsha, Hunan 410081, People's Republic
of China}

\begin{abstract}
\baselineskip=0.6 cm
\begin{center}
{\bf Abstract}
\end{center}

We present firstly the equation of motion for a test scalar particle coupling to Einstein tensor in the Schwarzschild-Melvin black hole spacetime through the short-wave approximation. Through analysing Poincar\'{e} sections, the power spectrum, the fast Lyapunov exponent indicator and the bifurcation diagram, we investigate the effects of coupling parameter on the chaotic behavior of the particles.
With the increase of the coupling strength, we find that the motion of coupled particle for the chosen parameters  becomes more regular and order for the negative couple constant. While, for the positive one,  the motion of coupled particles first undergoes a series of transitions between chaotic motion and regular motion and then falls into horizon or escapes to  spatial infinite.
Our results show that the coupling  brings about richer effects for the motion of the particles.

\end{abstract}
\maketitle

\newpage
\section{Introduction}

Chaos is a kind of non-periodic motion occurred in the non-linearly and deterministically dynamical systems, which is very sensitive to the initial conditions and presents the intrinsic random in systems \cite{Sprott,Ott,Brown1}. In general, it is very difficult to make a long-term prediction for chaotic motions, which means that chaotic systems possess many novel properties not shared by the usual dynamical systems and the non-linear interactions bring about richer physics. This triggers more attention for the study of chaotic dynamics in various physical fields. In general relativity,  we know that chaos does not emerge in the geodesic motion of the particle in the generic Kerr-Newman black hole spacetime because the equation of motion is variable-separable and the dynamical system is integrable \cite{Carter}. To ensure that the dynamical system describing motion of the particle is non-integrable and study the corresponding chaotic phenomenon, we have to resort to some spacetimes with complicated geometries or introduce some extra interactions. In this spirit, the chaotic trajectories of particle  have been found in multi-black hole spacetimes \cite{Cornish,Hanan}, or in the perturbed Schwarzschild spacetime \cite{Bombelli,Bombelli1,Bombelli2,Bombelli3}, or in the spacetime of a black hole immersed in magnetic field \cite{Karas}, or in the non-standard Kerr black hole spacetime described by Manko-Novikov metric \cite{Contopoulos,Contopoulos1,Contopoulos2,Contopoulos3,Hanwb2}, or in the accelerating and rotating black hole spacetime \cite{sbch}. Moreover, the chaotic phenomenon has also been investigated for charged particles moving in a magnetic field interacting with gravitational waves \cite{HD}, or for the pinning particles in Kerr spacetime \cite{Hanwb1}. More interestingly, after introducing ring strings instead of point particles, chaos in ring string dynamics has been found in the asymptotically flat Schwarzschild black hole spacetime \cite{Frolov}, in AdS-Schwarzschild black hole \cite{Zayas} and in AdS-Gauss-Bonnet black hole spacetimes \cite{MDZ}.

 One of the most interesting interactions is the coupling between scalar field and Einstein tensor described by the term $G^{\mu\nu} \partial_{\mu} \psi \partial_{\nu} \psi$.
 Such kinds of interactions were originally introduced by Horndeski \cite{ou18} and subsequently have been investigated extensively in cosmology to explain the cosmic accelerating expansion confirmed by the current observations \cite{ouz5,ouz6,ouz7,ouz8}.
 It is shown that the model of scalar field coupling to Einstein tensor has a ``good" dynamical property since that  equation of motion for the scalar field in this coupling theory is still a second-order differential equation. Moreover, the attractive features of such theory are that the derivative coupling term with Einstein tensor cannot only explain the accelerating expansion of the current Universe, but also solve naturally the problem of a graceful exit from inflation without any fine-tuned potential in the early Universe \cite{ouyz22,ouyz}. From the point of view of physics, a good theoretical model in cosmology should be examined by black hole physics since black hole
is another fascinating aspect in the modern theories of gravity.
Therefore, it is very necessary to study such coupling theory in black
hole physics. We \cite{ouhd} studied the dynamical evolution for a
scalar field coupled to Einstein¡¯s tensor in the background of
Reissner-Nordstr\"{o}m black hole spacetime and found a distinct
behavior of the coupled scalar field that the growing modes appear
as the coupling constant is larger than a certain threshold value.
Moreover, Minamitsuji \cite{ou15} investigated such coupling
theoretical model in braneworlds
 and the scalar-tensor theory, and found that a new phase transition from a Reissner-Nordstr\"{o}m
black hole to a hairy black hole takes place in asymptotically flat
spacetime because the Abelian $U(1)$ gauge symmetry is broken in the
vicinity of the black hole horizon when the coupling constant is
large enough \cite{ou17}. Other regular black hole solutions are also found in \cite{ou171,ou172}, which implies that such a coupling could contribute to the effective cosmological constant at infinity. These novel results have attracted more attention for studying the full properties for such special coupling model in black hole physics.

However, the above investigation focused mainly on the effects of such kind of coupling on the wave dynamical behavior of the coupled field.
It is natural to ask what effects of these couplings have on the motion of the particles in the background of a black hole. In order to reach this purpose, we must firstly obtain the equation of motion of the particle in a spacetime.
Considering a perturbational material field, one should get the equation of motion for a test particle from the corresponding wave dynamical equation by the short-wave approximation. In this way, the equation of motion of the coupled photon is obtained from the corrected Maxwell equation with the coupling between electromagnetic field and curvature tensors \cite{ou338,ou339,ou340,ou341,ou3}. With this technique, we here assume the coupled scalar field as a perturbation and obtain the equation of motion of the coupled scalar particle as a test particle, and then study such coupling effects on the motion of the particle in the background of a Schwarzschild-Melvin black hole \cite{Ernst}. The main reason why we select this spacetime as a background is that it describes the gravity of a black hole immersed in an external magnetic field, which is of some astrophysical interest because such kinds of the magnetic fields can be likely generated by currents in the accretion disk around a black hole in real astrophysical situations, especially at the center of galaxies. Moreover, its Einstein tensor has non-zero components and there exist chaotic phenomena in the geodesic motion of the particles in this background with an external magnetic field \cite{Karas} so that we can study such coupling effects on the chaotic motion of a test scalar particle.

The paper is organized as follows. In section II, we obtain the geodesic equation of a test scalar particle coupled to the Einstein tensor in a Schwarzschild-Melvin black hole by the short-wave approximation. In section III, we investigate the chaotic phenomenon in the motion of the particle coupled to Einstein tensor in the Schwarzschild-Melvin spacetime by the fast Lyapunov indicator, power spectrum, Poincar$\acute{e}$ section and bifurcation diagram. We probe the effects of this coupling  together with magnetic field on the chaotic behavior of a coupled scalar particle. Finally, we end the paper with a summary.

\section{Equation of motion for scalar particles coupling to the Einstein tensor in the Schwarzschild-Melvin black hole spacetime}

In this section, we will derive the equations of motion for a test particle coupled to the Einstein tensor in the Schwarzschild-Melvin black hole spacetime. The action describing scalar fields coupled to the Einstein tensor in the Einstein-Maxwell theory can be expressed as
\begin{equation}
\label{zyl}
S=\int d^4x\sqrt{-g}\bigg[\frac{R}{16\pi G}-\frac{1}{4}F_{\mu\nu}F^{\mu\nu}+\frac{1}{2}g^{\mu\nu}\partial_{\mu}\psi\partial_{\nu}\psi
+\frac{\alpha}{2}G^{\mu\nu}\partial_{\mu}\psi\partial_{\nu}\psi+\mu^2\psi^2\bigg],
\end{equation}
where the term $G^{\mu\nu}\partial_{\mu}\psi\partial_{\nu}\psi$ represents the coupling between the Einstein tensor $G^{\mu\nu}$ and the scalar field $\psi$ with mass $\mu$. The factor $\alpha$ is a coupling parameter with dimensions of length squared. The quantity $F_{\mu\nu}$ is the usual electromagnetic tensor. In order to study the motion of a test scalar particle in a background spacetime, as in the previous discussion, we treat scalar field $\psi$ as a perturbation field and then corresponding scalar particle as a test particle. After doing so, the coupling between scalar field and Einstein tensor does not modify the background spacetime and then the action (\ref{zyl}) admits a solution with the metric \cite{Ernst}
\begin{eqnarray}
\label{xy}
ds^{2} = \Sigma^{2}(-\Delta^{2} dt^{2}+\Delta^{-2} dr^{2}+r^{2} d\theta^{2})+\Sigma^{-2}r^{2}\sin^{2}\theta d\phi^{2},
\end{eqnarray}
which describes the gravity of a static black hole immersed in the extra magnetic field. The vector potential for the magnetic field is given by
\begin{eqnarray}
A_{\mu}dx^{\mu}=-\frac{Br^2\sin^2\theta}{\Sigma}d\phi.
\end{eqnarray}
Here $\Sigma$ and $\Delta$ are given by
\begin{eqnarray}
\label{xgm}
\Sigma = 1+\frac{1}{4}B^{2}r^{2}\sin^{2}\theta,\;\;\;\;\;\;\;\;\;\;\;\;\;\;\;\;\;\;\;\;
\Delta^{2}=1-\frac{2m}{r},
\end{eqnarray}
where $B$ is magnetic induction intensity of the extra magnetic field and $m$ is the mass of the black hole. As in the Schwarzschild case, the event horizon is located at $r=2m$ and there is no singularity outside the horizon. However, the polar circumference for the event horizon increases with the magnetic field, while the equatorial circumference decreases \cite{surf}.  Moreover,  due to the presence of the magnetic field, the Schwarzschild-Melvin spacetime is not asymptotically flat. The non-zero components of the Einstein tensor are
\begin{eqnarray}
\label{GG1}
&&G^{00}=\frac{4096B^{2}(r-2m\sin^{2}\theta)}{(2m-r)(
4+B^{2}r^{2}\sin^{2}\theta)^{6}},\;\;\;\;\;\;\;\;\;\;\;\;\;\;\;\;\;\;\;\;\;
G^{11}=\frac{4096B^{2}(r-2m)[r\cos^2\theta-(r-2m)\sin^{2}\theta]}{r^{2}(4
+B^{2}r^{2}\sin^{2}\theta)^{6}},\;\;\;\;\;\;\;\;\;\;\nonumber\\
&&G^{22}=-\frac{4096B^2[r\cos^2\theta-(r-2m)\sin^{2}\theta]}{
r^3(4+B^{2}r^{2}\sin^{2}\theta)^6},\;\;\;\;\;\;\;\;\;\;\;\;\;\;\;\;\;\;\;\;\;\;\;\;\;\;
G^{33}=\frac{16B^{2}(2m\sin^{2}\theta-r)}{r^{3}\sin^{2}\theta(4+B^{2}r^{2}\sin^{2}\theta)^{2}},\;\;\;\;\;\;\;\;\;\;\;\;\;\;\;\;\;\;\;\;\;\;\;\;\;\;\;\nonumber\\
&&G^{12}=G^{21}=\frac{4096B^{2}(2m-r)\sin2\theta}{r^{2}(4+B^{2}r^{2}\sin^{2}\theta)^{6}}.
\end{eqnarray}
Obviously, the Einstein tensor for this metric depends on  electromagnetic field. Thus, the coupling between scalar field and Einstein tensor in Schwarzschild-Melvin black hole spacetime can be really understand as a kind of interaction between scalar field and electromagnetic field. Theories containing the interaction between scalar field and electromagnetic field have been studied extensively, such as Einstein-Maxwell-Dilaton theory.

Varying the action (\ref{zyl}) with respect to $\psi$, one can find that the wave equation of scalar fields is modified,
\begin{eqnarray}
\frac{1}{\sqrt{-g}} \partial_{\mu} \bigg[\sqrt{-g}(g^{\mu\nu}+\alpha G^{\mu\nu}) \partial_{\nu} \psi\bigg]-\mu^2\psi = 0. \label{Eqsc2}
\end{eqnarray}
In order to get equation of the motion of a test scalar particle from the above
corrected Klein-Gordon equation (\ref{Eqsc2}), one can adopt the short-wave approximation in which the wavelength of the scalar particle
$\lambda_s$ is much smaller than the typical curvature scale $L$.
This means that the propagation process of the scalar particle does not
change the background gravitational field. The similar treatment for electromagnetic field are taken in \cite{ou338,ou339,ou340,ou341,ou3}
to probe the effect of the polarization direction on the propagation of photon in various background spacetime.
With this approximation, the scalar field $\psi$ can be rewritten as
\begin{eqnarray}
\label{bo}
\psi = f e^{iS}.
\end{eqnarray}
where $f$ is a real and slowly varying amplitude and $S$ is a rapidly varying phase. This means that the derivative term $f_{;\mu}$ is not dominated and can be neglected in this approximation. Moreover, the wave vector $k_{\mu}=\frac{\partial S}{\partial x^{\mu}}$ can be treated as the usual momentum $p_{\mu}$ in the theory of scalar particle. Inserting Eq.(\ref{bo}) into the action (\ref{Eqsc2}), we can find that the corrected Klein-Gordon equation (\ref{Eqsc2}) can be simplified as
\begin{eqnarray}
(g^{\mu\nu}+\alpha G^{\mu\nu}) k_{\mu}k_{\nu}=-1.\label{KG1}
\end{eqnarray}
Here, we set the mass of the scalar particle $\mu=1$.
It is obvious that the motion of the coupled particle is non-geodesic in the original metric.  Actually, these coupled scalar test  particles follow the geodesics in another effective spacetime with metric $\tilde{g}_{\mu\nu}$, which is related to the original metric $g_{\mu\nu}$ by
\begin{eqnarray}
\label{yxdg}
\tilde{g}^{\mu\nu}=g^{\mu\nu}+\alpha G^{\mu\nu}.
\end{eqnarray}
 To get a Hamiltonian formulation like in Ref.\cite{hamin}, one can introduce the momenta $p_{\mu}=\tilde{g}_{\mu\nu}\frac{dx^{\nu}}{d\lambda}$ ( where $\lambda$ is affine parameter ), and then obtain the Hamiltonian
\begin{eqnarray}
\label{Lagr1}
\mathcal{H} = \frac{1}{2}(g^{\mu\nu}+\alpha G^{\mu\nu}) k_{\mu}k_{\nu}= \frac{1}{2}\tilde{g}^{\mu\nu}k_{\mu}k_{\nu}.
\end{eqnarray}
With the effective metric $\tilde{g}_{\mu\nu}$ and the Hamiltonian equation, the
canonical equations of the coupled scalar particles can be written as
\begin{eqnarray}
\label{eqsm0}
\frac{d}{d\lambda}\bigg(\tilde{g}_{\mu\tau} \frac{dx^{\tau}}{d\lambda}\bigg)-\frac{1}{2}\tilde{g}_{\nu\tau;\mu}\frac{dx^{\nu}}{d\lambda} \frac{dx^{\tau}}{d\lambda}=0.
\end{eqnarray}
Correspondingly, Eq. (\ref{KG1}) becomes
\begin{eqnarray}\label{12}
\tilde{g}_{\mu\nu} \frac{dx^{\mu}}{d\lambda} \frac{dx^{\nu}}{d\lambda}=-1.
\end{eqnarray}
It means that the non-geodesic motion of the coupled particle in the original spacetime can be treated as a geodesic in the effective spacetime with metric $\tilde{g}_{\mu\nu}$. Here, we must point out that although Eqs (\ref{eqsm0}) with (\ref{12}) look just like the geodesic motion of the usual massive particles, the effective metric coefficient $\tilde{g}_{\mu\nu}$ depends on the types of particles and coupling, which means that the motions are actually different for various particles with different couplings.

Inserting Eqs.(\ref{xy}) and (\ref{GG1}) into (\ref{yxdg}) one can obtain the effective metric $\tilde{g}^{\mu\nu}$ for the scalar particle coupled to the Einstein tensor
\begin{eqnarray}
\label{gg1}
&&\tilde{g}^{00}=\frac{16 [256 \alpha  B^2 (r-2m\sin^{2}\theta)+r (4+B^{2}r^{2}\sin^{2}\theta)^4]}{(2 m-r)(4+B^{2}r^{2}\sin^{2}\theta)^6},\nonumber\\
&&\tilde{g}^{11}=\frac{16 [r (r-2 m) (4+B^{2}r^{2}\sin^{2}\theta)^4+256 \alpha  B^2 (r-2m) (2m\sin^{2}\theta+r\cos2\theta)]}{r^2(4+B^{2}r^{2}\sin^{2}\theta)^6},\nonumber\\
&&\tilde{g}^{22}=\frac{16 [r (4+B^{2}r^{2}\sin^{2}\theta)^4-256 \alpha  B^2 (r\cos2\theta+2m\sin^{2}\theta)]}{r^3(4+B^{2}r^{2}\sin^{2}\theta)^6},\nonumber\\
&&\tilde{g}^{33}=\frac{16 \alpha  B^2 (2 m\sin^2\theta-r)}{r^3\sin^2\theta(4+B^{2}r^{2}\sin^{2}\theta)^2}+\frac{(B^2 r^2\sin^2\theta+4)^2}{16r^{2}\sin^2\theta},\nonumber\\
&&\tilde{g}^{12}=\tilde{g}^{21}=\frac{4096 \alpha  B^2 \sin2\theta (2 m-r)}{r^2 (4+B^{2}r^{2}\sin^{2}\theta)^6}.
\end{eqnarray}
The presence of $\tilde{g}^{12}$ and $\tilde{g}^{21}$ shows that the motion of coupled the scalar particle becomes more complex than that of in the case without the coupling.
It is obvious that the coefficients of the above effective metric are independent of the coordinates $t$ and $\phi$, which yields two constants of motion of the geodesics, i.e., the energy and the $z$-component of the angular momentum
\begin{eqnarray}
\label{EL}
E=-\frac{\partial
\mathcal{L}}{\partial\dot{t}}=-\tilde{g}_{00}\dot{t},\;\;\;\;\;\;\;\;\;\;\;\;\;\;
L_{z}=\frac{\partial
\mathcal{L}}{\partial\dot{\varphi}}=\tilde{g}_{33}\dot{\varphi},
\end{eqnarray}
which depend also on the coupling between scalar particle and Einstein tensor since $\tilde{g}_{00}$ and $\tilde{g}_{33}$ are functions of coupling parameter $\alpha$.
Substituting these two conserved quantities into Eqs.(\ref{eqsm0})-(\ref{12}), one can obtain the equation of motion with a constraint condition $\Delta H=0$, where
\begin{eqnarray}
\label{ydfc1}
\Delta H\equiv\frac{E^{2}}{\tilde{g}_{00}}+\tilde{g}_{11}\dot{r}^{2}
+\tilde{g}_{22}\dot{\theta}^{2}+\frac{L^{2}}{\tilde{g}_{33}}+2\tilde{g}_{12}\dot{r}\dot{\theta}+1.
\end{eqnarray}
for a massive scalar particle coupling to the Einstein tensor in the Schwarzschild-Melvin black hole spacetime. As in the case without coupling, these equations of motion cannot be variable-separable, which implies that the motion of a particle coupled to Einstein tensor could be chaotic in the Schwarzschild-Melvin black hole spacetime. In the next section, we will investigate the effect of the coupling on the chaotic behavior of the coupled scalar particle.

\section{The chaotic motion of scalar particles coupling to the Einstein tensor in the Schwarzschild-Melvin black hole spacetime}

In order to study the chaotic motion of scalar particles coupled to Einstein's tensor, we must solve the differential equations (\ref{eqsm0}) and (\ref{12}) numerically with high precision. The main reason is that the
motion of a particle in the chaotic region is very sensitive to initial value and the larger numerical errors may produce pseudo chaos, which is not the real motion of the particle. Here, we
adopt to the corrected fifth-order Runge-Kutta method suggested in
Refs.\cite{dz42,dz43,dz44}, in which the velocities $(\dot{r}, \dot{\theta})$ are corrected in integration and the numerical deviation is
pulled back in a least-squares shortest path. The scale factor of
the velocity correction $\xi$ for a scalar particle coupled to Einstein's tensor in the Schwarzschild-Melvin black hole spacetime (\ref{xy}) is
\begin{eqnarray}
\label{lmd}
\xi=\sqrt{-\frac{1 +E^{2}/\tilde{g}_{00}+L^{2}/\tilde{g}_{33}}{\tilde{g}_{11}\dot{r}^{2}+\tilde{g}_{22}\dot{\theta}^{2}
+2\tilde{g}_{12}\dot{r}\dot{\theta}}}.
\end{eqnarray}
\begin{figure}[ht]
\center{\includegraphics[width=7cm ]{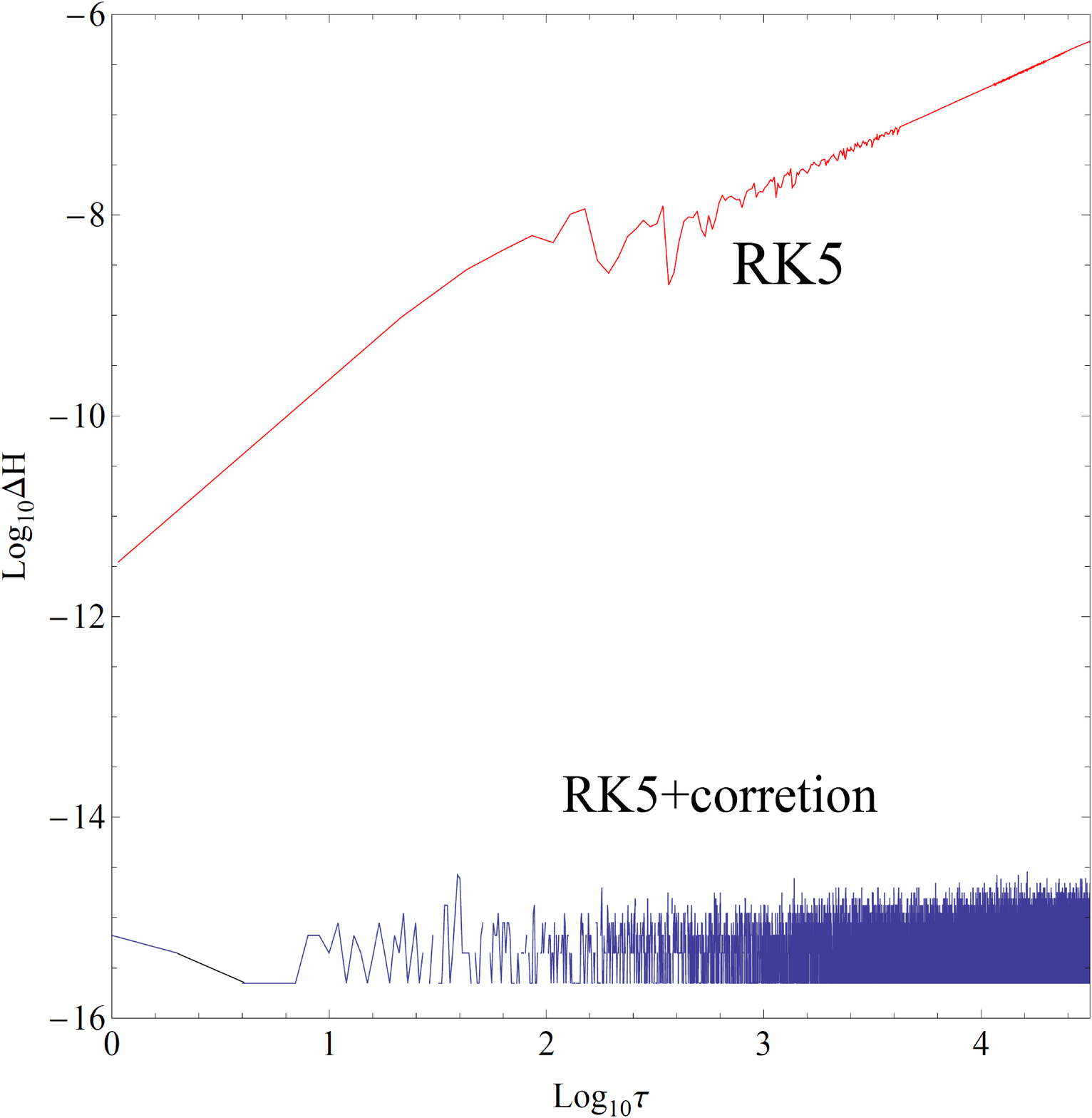}
\caption{The energy errors $\Delta H$ with time computed by RK5 and the velocity correction method (RK5+Correction) in the Schwarzschild-Melvin black hole spacetime.}
\label{ffdb1}}
\end{figure}
As in Refs.\cite{dz42,dz43,dz44}, we find that with the velocity-correction fifth-order Runge-Kutta method the numerical error is controlled greatly and the value of $\Delta H$ is kept below $10^{-14}$, which is shown in Fig.(1). This means that this correction method is so powerful that it can avoid the pseudo chaos caused by numerical errors.

It is well known that the motion of the particle is determined by the parameters and initial conditions of the dynamical system. As in usual cases, we set the mass of black hole $m=1$. The choice of magnetic field parameter is based on a consideration that the magnetic induction intensity around black hole satisfies $B\ll  m^{-1}$ (in the natural unit $G=c=\hbar=1$) even for the very strong magnetic fields in centers of real galaxies \cite{Konoplya}. Moreover,  we find that it is more difficult to obtain the stable orbit for the particle in the cases with the high numerical value of $B$ in our numerical calculation. For the parameters and initial conditions of particle, it should be stressed that the choice should in principle be arbitrary. Here, we  take a set of parameters ($E=0.99$, $L=3.9$, $m=1$, $B=0.02$) and initial conditions ($r(0)=9.7$, $\theta(0)=\frac{\pi}{2},\dot{r}(0)=0$) only as an example to illustrate the change of the degree of disorder of particle orbits with the coupling parameter $\alpha$. The other sets of the parameters and initial conditions are also chosen to analyse the orbit of particles in the subsequent discussion.

\begin{figure}
\center{\includegraphics[width=15cm ]{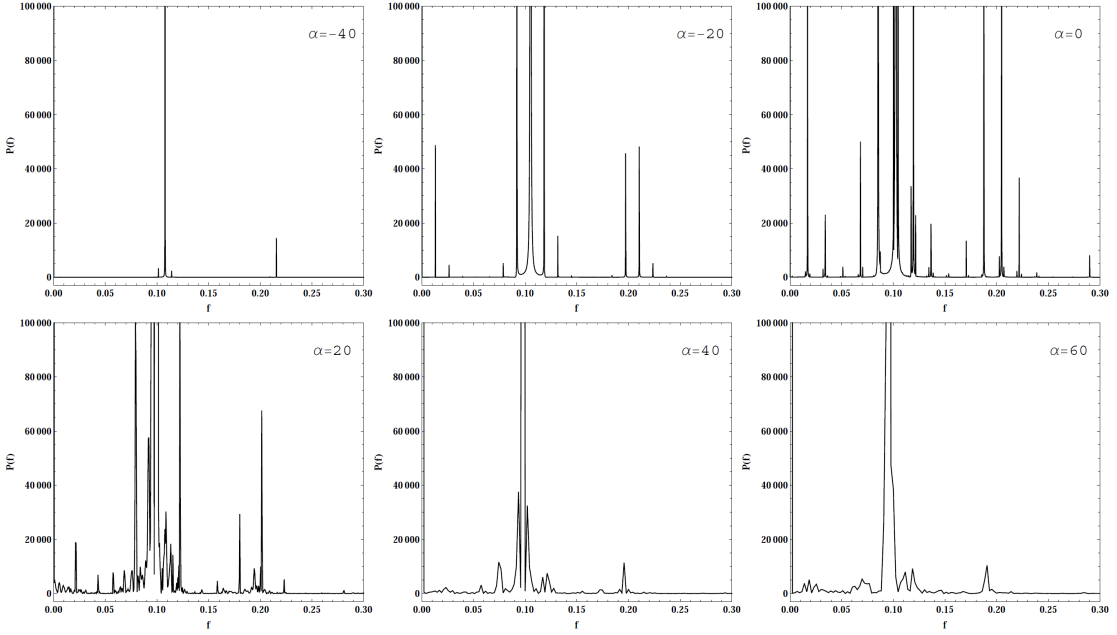}
\caption{The power spectrum of the solutions for different $\alpha$  with a set of parameters ($E=0.99$, $L=3.9$, $m=1$, $B=0.02$) and initial conditions ($r(0)=9.7$, $\theta(0)=\frac{\pi}{2},\dot{r}(0)=0$).}
\label{dlp}}
\end{figure}

Let us now to analyse the frequency components in these solutions by the power spectrum method in which the heights of the bars are dominated by the square of the amplitude related to each frequency in a Fourier
decomposition \cite{Sprott,Ott}. From Fig.(\ref{dlp}), we find that it is a discrete spectrum for $\alpha=0$, $-20$ and $-40$, which means that in the cases $\alpha=0$, $-20$ and $-40$ the motion of the particle is the order and regular motion. Compared with the case with $\alpha=0$, we also find that there exist much fewer frequency components in the cases with negative $\alpha$. For the signals with $\alpha=20$, $40$ and $60$, it is obvious that there exists a continuous spectrum, and then the corresponding motion is chaotic for the coupled particle.
 Thus, with the increase of the coupling strength, one can find that for the chosen parameters the degree of disorder and non-integrability of the motion of the particle increases  for the positive $\alpha$ and decrease for the negative one.

The Lyapunov indicator is another useful tool to discern the chaotic orbits of particles by measuring two adjacent orbits over time with the average separation index \cite{Ott,lci2}. The motion of the particle is chaotic if there exists a positive Lyapunov exponent and it is order if all of Lyapunov exponents are negative. Recently, a faster and more convenient Lyapunov indicator is proposed by Froeschl\'{e} \textit{et al} in \cite{ch47} and then it is extended to the cases in general relativity by Wu \cite{ch50}. In the curve spacetime, the version of the Fast Lyapunov indicator (FLI) with the two particles method can be expressed as
\begin{equation}
FLI_{k}(\tau)=-k\cdot(1+\log_{10}\|\Delta \mathbf{x}(0)\|)+\log_{10}\frac{\|\Delta \mathbf{x}(\tau)\|}{\|\Delta \mathbf{x}(0)\|}.
\end{equation}
\begin{figure}
\center{\includegraphics[width=15cm ]{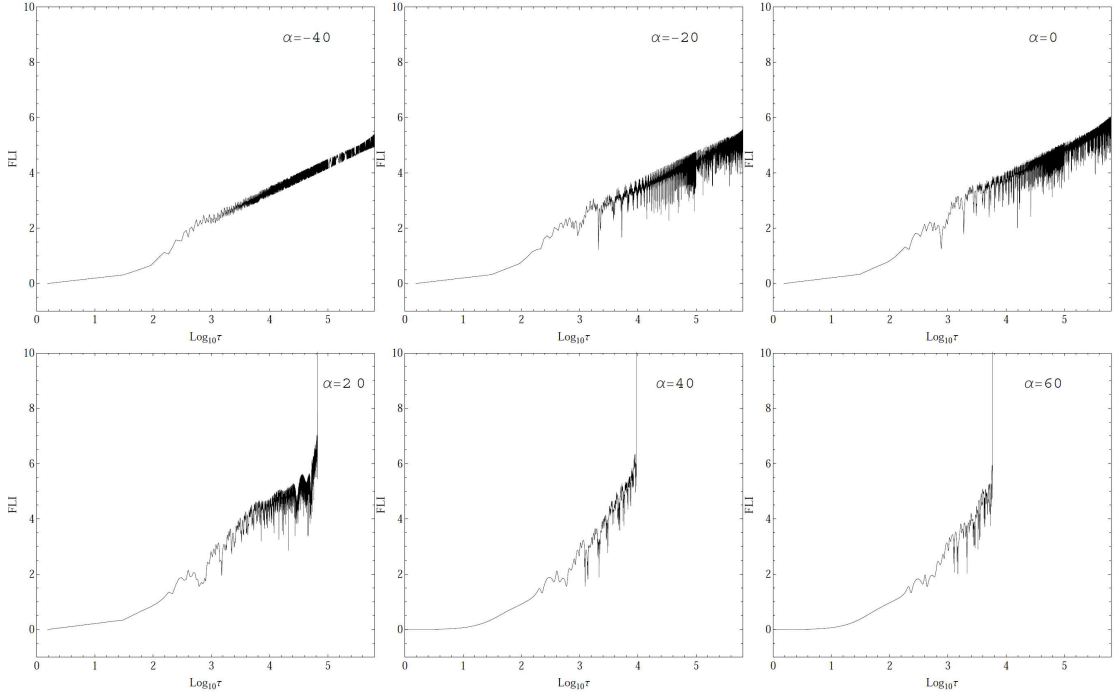}
\caption{The Fast Lyapunov indicators with two nearby trajectories for the solutions plotted in Fig.(2).}
\label{fli}}
\end{figure}
Here $\|\Delta \mathbf{x}(\tau)\|=\sqrt{|g_{\mu\nu}\Delta \mathbf{x}^{\mu}\Delta \mathbf{x}^{\nu}|}$, and  $\Delta \mathbf{x}^{\alpha}(\tau)$ is the deviation vector between two nearby trajectories at proper time $\tau$, which is given by
\begin{equation}
\Delta \mathbf{x}^{\alpha}(\tau)=\tilde{x}^{\alpha}(\tau)-x^{\alpha}(\tau).
\end{equation}
The quantity $k$ stands for the sequential number of renormalization which is used to avoid the numerical saturation arising from the fast separation of the two adjacent orbits. It is shown that FLI($\tau$) grows with exponential rate for chaotic motion, even for weak chaotic motion, and  grows algebraically with time for the regular resonant orbit and for the periodic one \cite{Ott,lci2,ch47,ch50}.
In Fig. (\ref{fli}), we plot the FLI with proper time for the solutions plotted in Fig.(2)  with $\Delta r(0)=10^{-8},\Delta \dot{r}(0)=0$ and $\Delta \theta(0)=0$. From Fig. (\ref{fli}), we find that with increase of time $\tau$, FLI($\tau$) grow with exponential rate for the signal with $\alpha=20$, $40$ and $60$, but
with polynomial rate in the cases with  $\alpha=0$, $-20$ and $-40$. This confirms further that in Fig.(2) the orbits in the cases with $\alpha=20$, $40$ and $60$ are chaotic and the orbits with $\alpha=0$, $-20$ and $-40$ are ordered in this case. It also supports that for chosen parameters and initial conditions the degree of disorder and non-integrability of the motion of the particle increases with the coupling parameter $\alpha$. Especially, the presence of $\alpha$ could mean that the motion of the particle changes from the order motion in the case without coupling to the chaotic motion, which implies that the coupling between scalar particle and Einstein tensor brings about richer properties for the motion of the particles.

\begin{figure}
\includegraphics[width=16cm ]{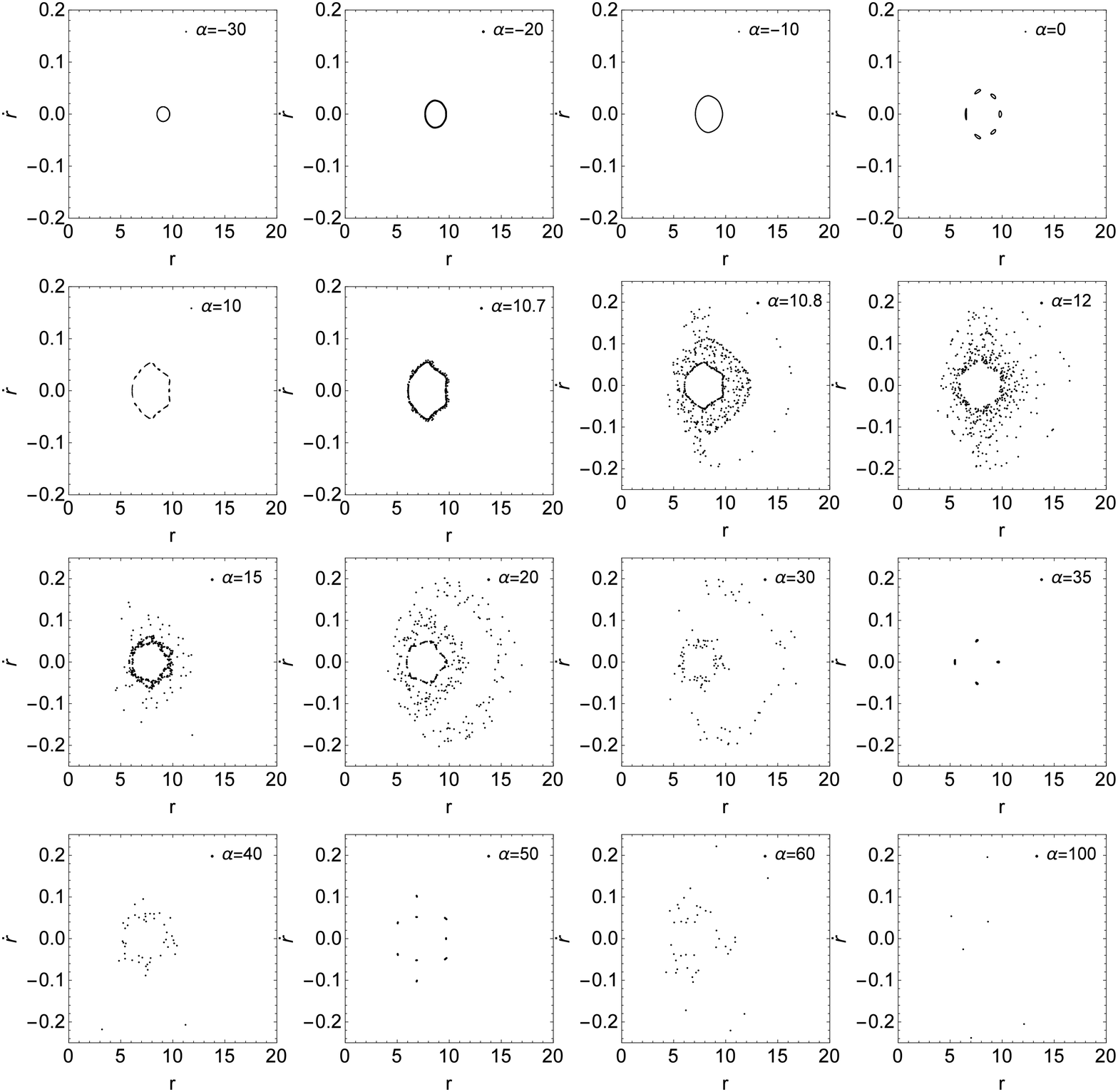}
\caption{The Poincar\'{e} surface of section ($\theta=\frac{\pi}{2}$) with different coupling parameter $\alpha$ and fixed $B=0.02$. The initial conditions are set as $r(0)=9.7$, $\theta(0)=\frac{\pi}{2}$ and $\dot{r}(0)=0$.}
\label{p1}
\end{figure}
\begin{figure}
\includegraphics[width=16cm ]{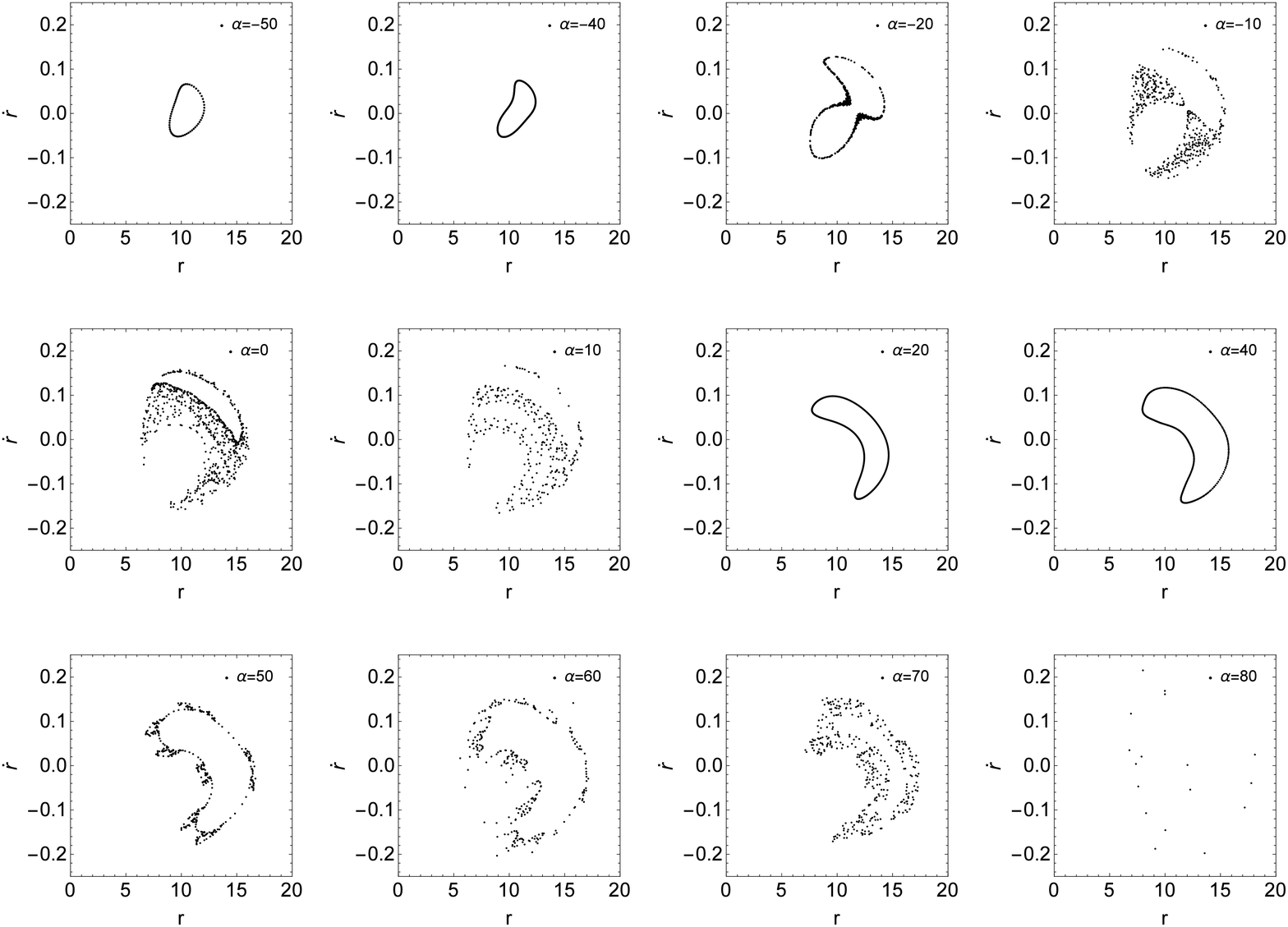}
\caption{The Poincar\'{e} surface of section ($\theta=\frac{2\pi}{5}$) with different coupling parameter $\alpha$ and fixed $B=0.02$. The initial conditions are set as $r(0)=12.0$, $\theta(0)=\frac{3\pi}{5}$ and $\dot{r}(0)=0$.}
\label{p2}
\end{figure}
\begin{figure}
\includegraphics[width=13cm ]{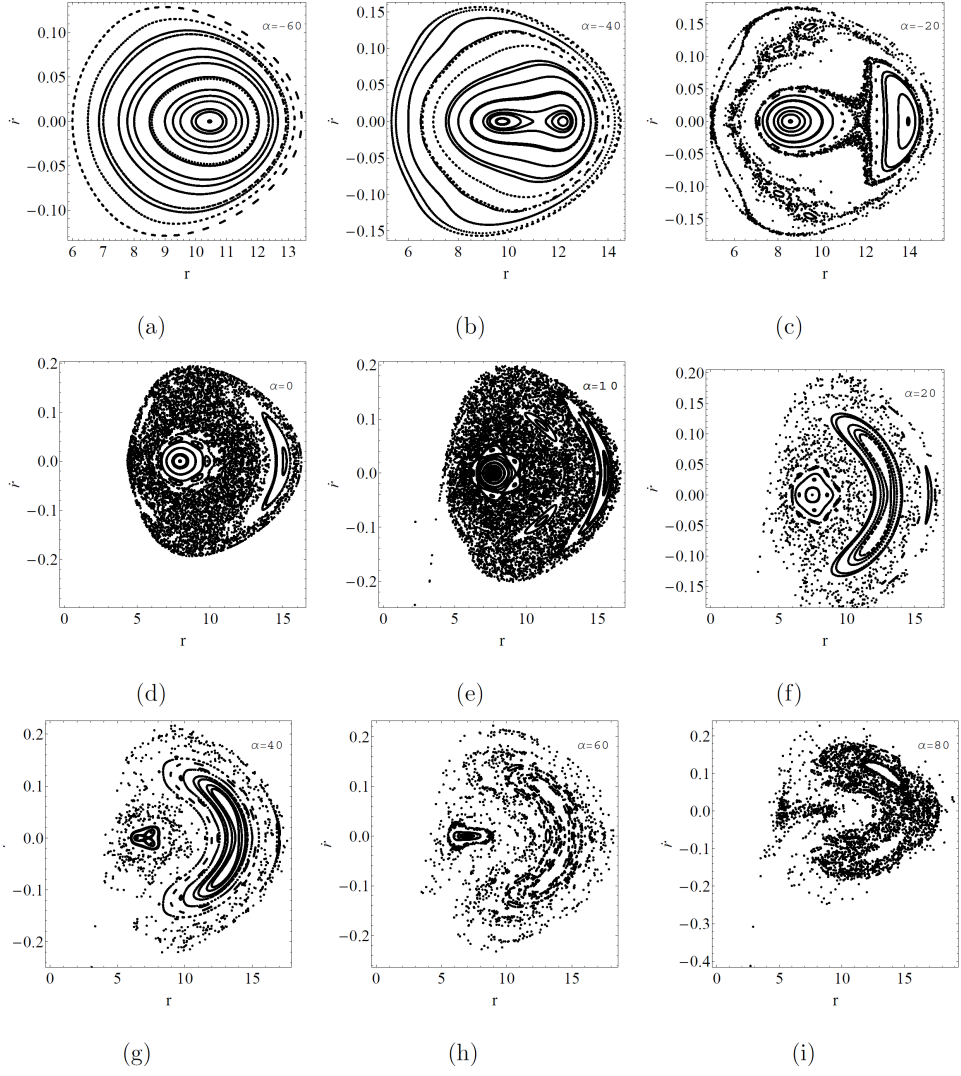}
\caption{The Poincar\'{e} surface of section ($\theta=\frac{\pi}{2}$) with different coupling parameter $\alpha$ and fixed $B=0.02$.}
\label{aaa}
\end{figure}
\begin{figure}
\includegraphics[width=13cm ]{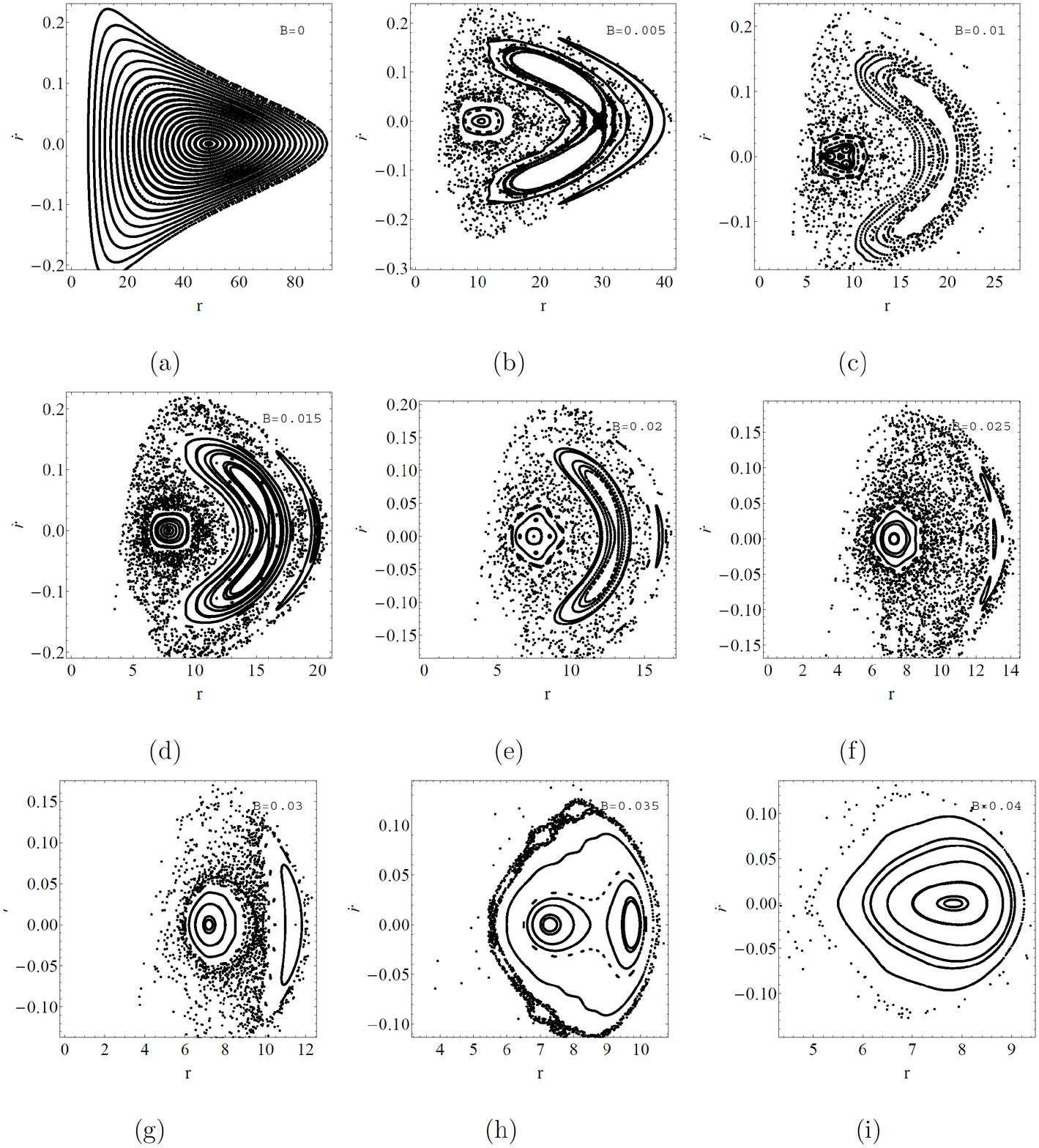}
\caption{The Poincar\'{e} surface of section ($\theta=\frac{\pi}{2}$) with different magnetic induction $B$ and fixed coupling parameter $\alpha=20$.}
\label{bbb}
\end{figure}

Poincar\'{e} section is an intersection of trajectory of a
continuous dynamical system with a given hypersurface which is
transversal to the trajectory in the phase space. It can be applied to discern chaotic motions of particles moving in the curve spacetime. In general, the
solutions of the continuous dynamical system with different initial conditions can be classified as three kinds by the intersection points in a Poincar\'{e} section. The periodic solutions and the quasi-periodic correspond to a finite
number of points and a series of close curves in the Poincar\'{e} section, respectively. The chaos solutions correspond to strange patterns of dispersed points with complex boundaries \cite{an25}.
In Fig.(\ref{p1}), Poincar\'{e} sections with $\theta=\frac{\pi}{2}$ on the plane $(r, \dot{r})$  for different coupling parameters $\alpha$ are plotted for the motion of a coupled particle in the the Schwarzschild-Melvin black hole spacetime with the fixed parameters $B=0.02$, $E=0.99$, $L=3.9$ and the initial conditions $\{$ $r(0)=9.7$; $\dot{r}(0)=0$; $\theta(0)
=\frac{\pi}{2}$$\}$. We find that for $ \alpha<10.7$ the phase path
is a quasi-periodic Kolmogorov-Arnold-Moser (KAM) tori and the behavior of this system is non-chaotic.  Especially, there are more complicated KAM tori trajectories for $\alpha=0$. It is composed of six  secondary KAM tori belonging to the same trajectories where the successive points jump from one loop to the next. These little loops are called a chain of islands. With the coupling parameter $\alpha$ increasing, the chain of islands are joined together and become a big KAM tori. This shows that trajectory is regular and integrable in this case. However, when $\alpha\sim(10.7, 30)$, we find that KAM tori is destroyed and the corresponding trajectory is non-integrable, which indicates that the behavior of this system is chaotic. As the value of $\alpha\sim (31,37)$, one can find that the motion of the particle becomes order again. With the further increasing of $\alpha$, the motion of the particle undergoes a series of transitions between chaotic motion and regular motion. As $\alpha\simeq100$,
the particle falls finally into the event horizon of the black hole or spatial infinite after undergoing some chaotic oscillations around black hole. In this case, one cannote that there exist a few discrete points in the Poincar\'{e} section, but it is different essentially from those in the case of usual multiple-periodic motion. Thus, with the increase of the coupling strength, one can find that for the chosen parameters the degree of disorder and non-integrability of the motion of the particle almost increases  for the positive $\alpha$ and almost decrease for the negative one. It is shown in Figs.(\ref{dlp}) and (\ref{fli})  that the motion is regular when $\alpha=0$ for the selected initial conditions set $\{$ $r(0)=9.7$; $\dot{r}(0)=0$; $\theta(0)=\frac{\pi}{2}$$\}$. It does not means that the motion is regular in the case $\alpha=0$ with any initial condition set since the behavior of non-linear dynamical system depends on the choice of the initial conditions.
In ref.\cite{Karas}, the chaotic motion of a test particle is found in the case $\alpha=0$ for the proper initial condition. Here, we adopt another initial condition $\{$ $r(0)=12.0$; $\dot{r}(0)=0$; $\theta(0)=\frac{3\pi}{5}$$\}$ as an example to indicate that the chaotic motion is allowable in the case $\alpha=0$, which is  shown in Fig.(\ref{p2}). We also investigate the dependence of
Poincar\'{e} sections with $\theta=\frac{2\pi}{5}$ on the coupling parameters $\alpha$ with this initial condition.  For the negative $\alpha$, we find that with the increase of the coupling intensity, the chaos becomes weak monotonously. As $\alpha\leq-25$, the KAM tori recovers and the motion is regular. For the positive $\alpha$, with the increase of the coupling intensity, one can find that the behavior of the system undergoes a process from chaotic to regular then to chaotic. Correspondingly,  the non-integrability of the motion of the particle in Fig.(\ref{p2}) first decreases and then increases with $\alpha$. Similarly, for the larger $\alpha$, we can obtain a kind of unstable escape solutions for this chosen initial condition as in the previous discussion. Thus, the dependence of the non-integrability of the motion on the coupling parameter $\alpha$ depends on the initial conditions and the parameters of system.

In Figs.(\ref{aaa}) and (\ref{bbb}), we also plot the Poincar\'{e} surface of section ($\theta=\frac{\pi}{2}$) on the plane $(r, \dot{r})$ for the motion of the scalar particle with different initial conditions.  From Fig.(\ref{aaa}), we find that the chaotic region decreases with the strength of the coupling between Einstein tensor and particle for the negative $\alpha$. When $ \alpha $ is positive, the chaotic region first decreases and then increases with the coupling strength. Moreover, it is shown in Fig.(\ref{bbb}) that for fixed $\alpha=20$ the chaotic region of this system  first increases and then decreases with the increase of the magnetic induction $B$. As $B=0$, one can find that only some closed curves appear in Poincar\'{e} surface of section, of which the motion is orderly and non-chaotic. The main reason is that in this case the Schwarzschild-Melvin black hole spacetime reduces to the usual Schwarzschild one, in which all of components of the Einstein tensor vanish and then the motions of the coupled particles are consistent with those of particles without coupling.

\begin{figure}
\center{\includegraphics[width=16.5cm ]{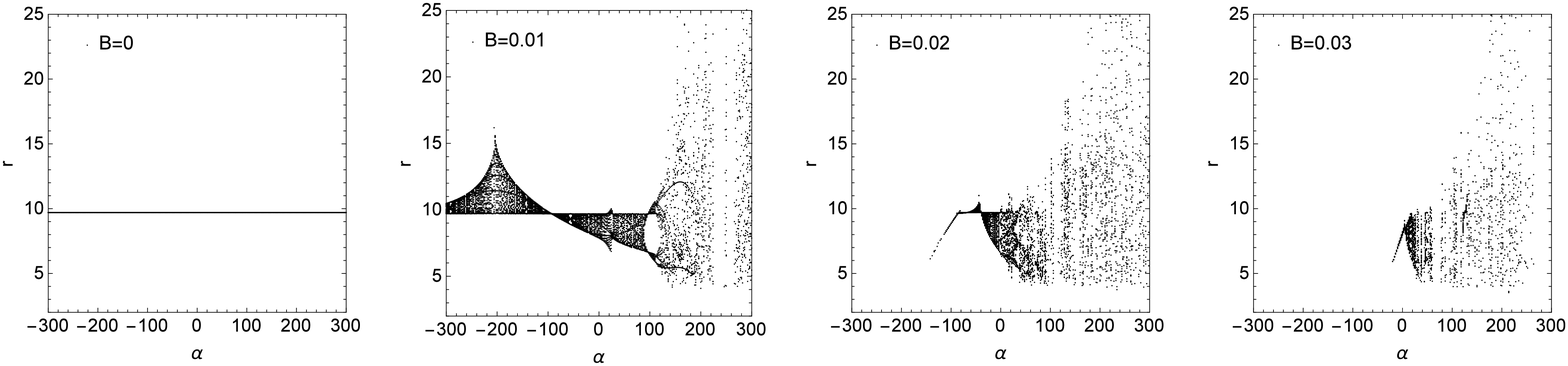}\\
\includegraphics[width=16.5cm]{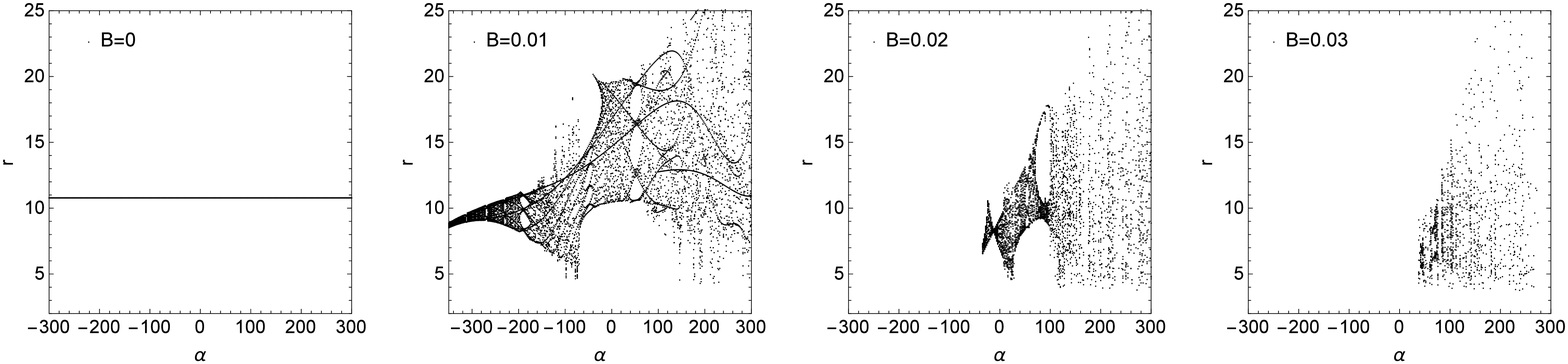}
\caption{The bifurcation diagram with coupling parameter $\alpha$ for different $B$ in the Schwarzschild-Melvin black hole spacetime. The set of initial conditions for the graphics are $\{$ $r(0)=9.7$;
$\dot{r}(0)=0$; $\theta(0) =\frac{\pi}{2}$$\}$ in the top row  and  are $\{$ $r(0)=12$;
$\dot{r}(0)=0.1$; $\theta(0) =\frac{2\pi}{5}$$\}$ in the bottom row.}
\label{fcaa1}}
\end{figure}

\begin{figure}
\center{\includegraphics[width=16.5cm]{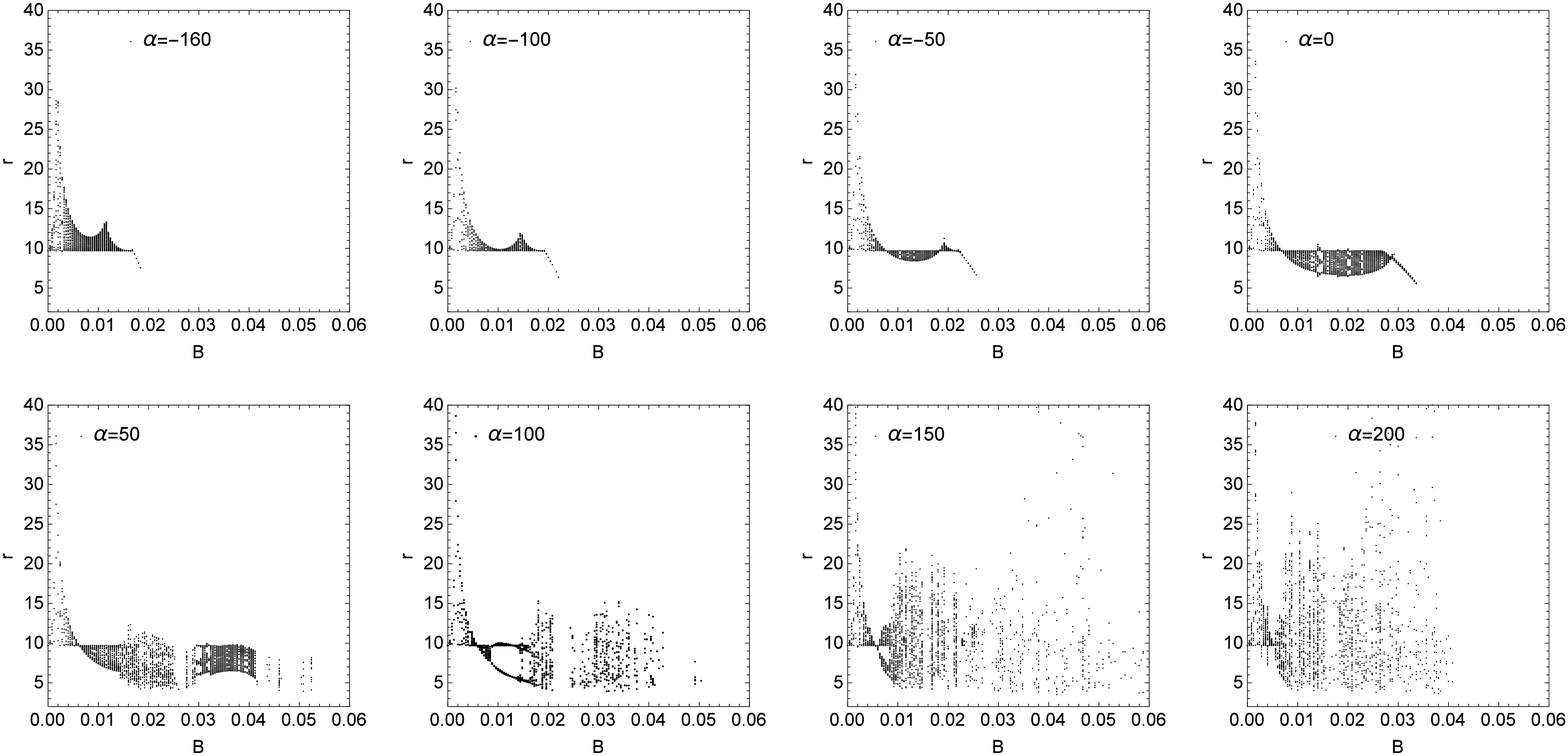}\\
\includegraphics[width=16.5cm]{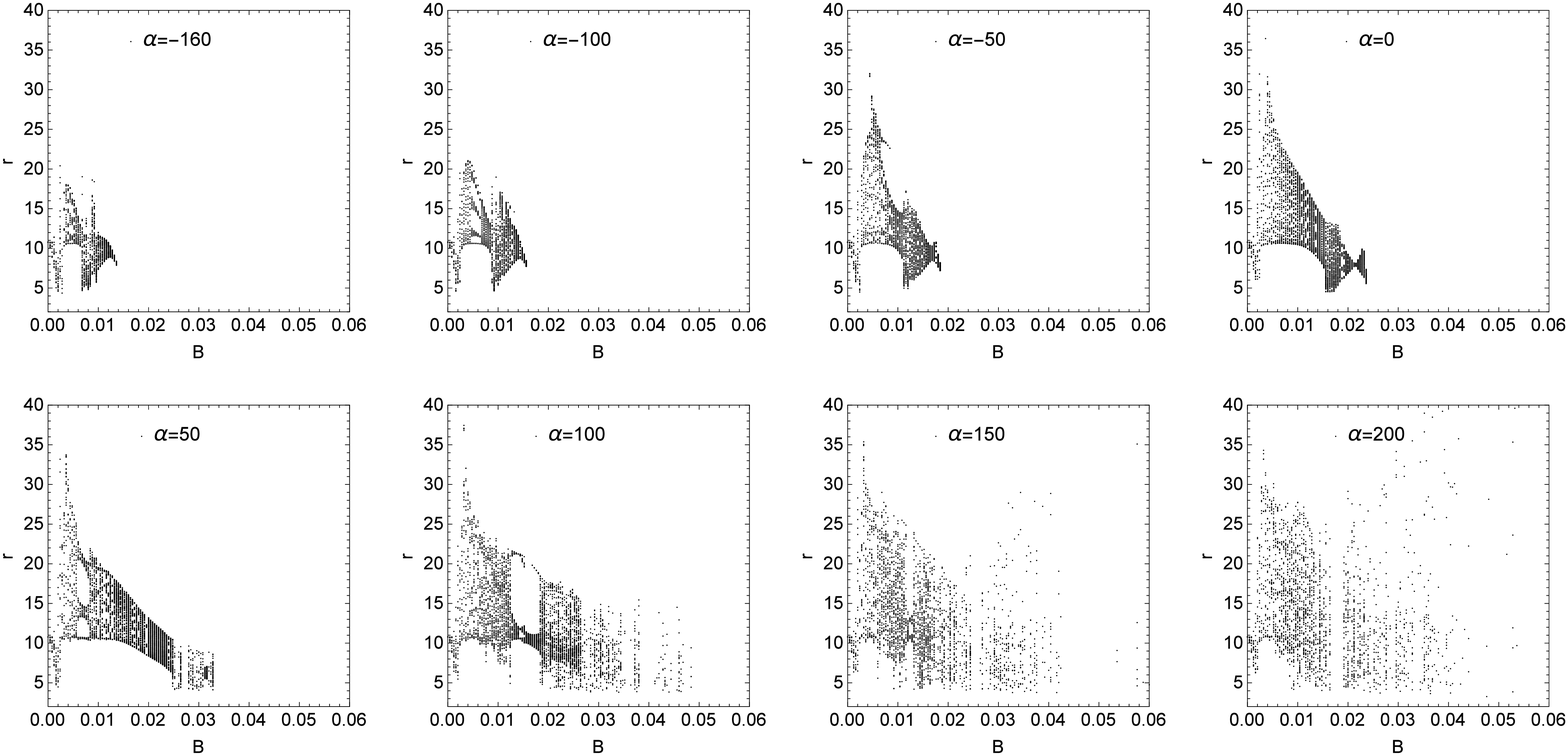}
\caption{The bifurcation diagram with magnetic field parameter $B$ for different $\alpha$ in the Schwarzschild-Melvin black hole spacetime. The set of initial conditions for the graphics are $\{$ $r(0)=9.7$;
$\dot{r}(0)=0$; $\theta(0) =\frac{\pi}{2}$$\}$ in the top two rows  and  are $\{$ $r(0)=12$;
$\dot{r}(0)=0.1$; $\theta(0) =\frac{2\pi}{5}$$\}$ in the bottom two rows.}
\label{fcbb1}}
\end{figure}

The bifurcation diagram can tell us about the dependence of the dynamical behaviors of system on the dynamical parameters, which could lead to the chaotic phenomenon in the system. In Figs.(\ref{fcaa1}) and (\ref{fcbb1}), we plot the bifurcation diagram of the radial coordinate $r$ with the coupling parameter $\alpha$ and magnetic induction $B$ for the coupled scalar particles moving in the Schwarzschild-Melvin black hole spacetime with fixed $E=0.99$ and $L=3.9$. Here we chose two set of initial conditions (i.e., $\{$ $r(0)=9.7$, $\dot{r}(0)=0$ and $\theta(0) =\frac{\pi}{2}$$\}$ and $\{$ $r(0)=12.0$, $\dot{r}(0)=0.1$ and $\theta(0) =\frac{2\pi}{5}$$\}$). As $B=0$, one can find that in both cases there is only a periodic solution and no bifurcation for the dynamical system, which means that the motions of particles are regular in this case. For the case with magnetic field, it is easy to find that there exist periodic, chaotic and escape solutions, which depend on the magnetic field parameter $B$ and the coupling parameter $\alpha$. Moreover, we can find that with the increase of the parameters $B$ and $\alpha$ the motion of coupled scalar particle transform among  single-periodic,  multi-periodic and chaotic motions, and the effects of the parameters $B$ and $\alpha$ on the motion of the coupled scalar particle are very complex, which are typical features of bifurcation diagram for the usual chaotic dynamical system. However, Figs.(\ref{fcaa1}) and (\ref{fcbb1}) tell us that the lower bound of $\alpha$ for the existence of the oscillation solution increases with the magnetic field parameter $B$. With the increase of the coupling strength, the upper bound of $B$ for the existence of the oscillation solution decreases with $\alpha$ for the negative $\alpha$, and almost increases for the positive one. These results imply that the scalar field coupling to the Einstein tensor yields  much richer effects on the motion of the particles.

\section{Summary}

In this paper we present firstly the equation of motion for a test scalar particle coupling to Einstein tensor in the Schwarzschild-Melvin black hole spacetime through the short-wave approximation. And then, we
have studied the dynamical behaviors of the motion of the test coupled particles by numerical method. Through analysing Poincar\'{e} sections,  power spectrum,  fast Lyapunov exponent indicator and  bifurcation diagram, we probe the effects of the coupling between scalar particle and Einstein tensor on the chaotic motion of the particles in the Schwarzschild-Melvin black hole spacetime. For fixed $B=0.02$, we find that
the chaotic region in Poincar¨¦ sections decreases monotonously with the strength of the coupling intense for the negative , but it first decreases and then increases in the case with the positive $\alpha$. For fixed $\alpha=20$,
one can find with the increase of magnetic induction $B$, the chaotic region of this system  first increases and then decreases.
The bifurcation diagram shows that the system undergoes a series process from regular to chaotic with the increase of $\alpha$. Moreover, we also find that the lower bound of $\alpha$ for the existence of the oscillation solution increases with the magnetic field parameter $B$. With the increase of the coupling strength, the upper bound of $B$ for the existence of the oscillation solution decreases with $\alpha$ for the negative $\alpha$, and almost increases for the positive one.  These results show that the coupling between scalar particle and Einstein tensor  yields richer effects on the motion of the particles in the Schwarzschild-Melvin black hole  spacetime. It would be of interest to generalize our study to the cases of the coupling with other curvature tensors, such as Weyl tensor etc. Work in this direction will be reported in the future.

\section{\bf Acknowledgments}

This work was partially supported by the National Natural
Science Foundation of China under Grant  No. 11475061,
the construct program of the National Key Discipline, and the Open
Project Program of State Key Laboratory of Theoretical Physics,
Institute of Theoretical Physics, Chinese Academy of Sciences, China
(No.Y5KF161CJ1).


\begin{thebibliography}{99}
\baselineskip=0.6cm



\bibitem{Sprott} J. C. Sprott, \textit{Chaos and Time-Series Analysis}, Oxford University Press, 2003.
\bibitem{Ott} E. Ott, \textit{Chaos in Dynamical Systems}, Cambridge University Press, Second Edition 2002.
\bibitem{Brown1}R. Brown and L. O. Chua,  Int. J.
Bifurcation and Chaos {\bf6},  219 (1996); Int. J. Bifurcation and Chaos {\bf8}, 1 (1998).
\bibitem{Carter}B. Carter, Phys. Rev. {\bf174} (5), 1559-1571 (1968).

\bibitem{Cornish} N. J. Cornish, C. P. Dettmann and N. E. Frankel, Phys. Rev. D {\bf50} (1994) R618-621,[arXiv:gr-qc/9402027].

\bibitem{Hanan}W. Hanan and E. Radu, Mod. Phys. Lett. {\bf A} 22 (2007) 399-406, [gr-qc/0610119].

\bibitem{Bombelli}L. Bombelli and E. Calzetta, Class. Quant. Grav. {\bf9}, 2573 (1992).
\bibitem{Bombelli1}J. M. Aguirregabiria, Phys. Lett. A {\bf224}, 234 (1997).
\bibitem{Bombelli2} Y. Sota, S. Suzuki and K. i. Maeda,  Class. Quant. Grav. {\bf13}, 1241 (1996).
\bibitem{Bombelli3}V. Witzany, O. Semerak and P. Sukova, Mon. Not. Roy. Astron. Soc. {\bf451} (2): 1770-1794 (2015).
\bibitem{Karas}V. Karas and D. Vokrouhlicky, Gen. Relativ. Gravit. {\bf24},729 (1992).
\bibitem{Contopoulos} J. R. Gair, C. Li, and I. Mandel, Phys. Rev. D {\bf77}, 024035 (2008).
\bibitem{Contopoulos1}G. Contopoulos, G. L. Gerakopoulos and T. A. Apostolatos, Int. J. Bifurc. Chaos {\bf21},  2261-2277  (2011); G. L. Gerakopoulos, G. Contopoulos and T. A. Apostolatos, arXiv:1408.4697.
\bibitem{Contopoulos2}F. L. Dubeibe, L. A. Pachon and J. D. Sanabria-Gomez, Phys. Rev. D {\bf 75}, 023008 (2007).
\bibitem{Contopoulos3}E. Gueron and P. S. Letelier, Phys. Rev. E {\bf66}, 046611 (2002).

\bibitem{Hanwb2} W. B. Han, Phys. Rev. D {\bf 77}, 123007 (2008).


\bibitem{sbch} S. Chen, M. Wang and J. Jing, J. High Energy Phys.{\bf09}, 082 (2016).

\bibitem{HD} H.Varvoglis, and D.Papadopoulos, Astronomy And Astrophysics {\bf261}, 664 (1992).

\bibitem{Hanwb1} W. B. Han, Gen Relativ Gravit {\bf 40}, 1831 (2008).


\bibitem{Frolov}A. V. Frolov and A. L. Larsen,  Class. Quant. Grav. {\bf16 }, 3717-3724 (1999).
\bibitem{Zayas} L. A. P. Zayas and C. A. Terrero-Escalante, J. High Energy Phys. {\bf09}, 094 (2010).
\bibitem{MDZ}D. Z. Ma, J. P. Wu and J. F. Zhang, Phys. Rev. D {\bf89}, 086011 (2014).








\bibitem{ou18}G. W. Horndeski, Int. J. Theor. Phys., {\bf10}, 363 (1974).

\bibitem{ouz5}A. G. Riess, \textit{et al}, Astrophys. J. {\bf 116}, 1009 (1998).
\bibitem{ouz6}P. Bernardis, \textit{et al}, Nature (London) {\bf 404}, 955 (2000).
\bibitem{ouz7}S. Perlmutter, Astrophys. J. {\bf 517}, 565 (1999).
\bibitem{ouz8}D. J. Eisenstein, Astrophys.J. {\bf 633}, 560 (2005).

\bibitem{ouyz22} S. V. Sushkov,  Phys. Rev. D {\bf 80}, 103505 (2009).
\bibitem{ouyz}C. J. Gao, J. Cosmo. Astro. Phys.{\bf 6}, 23 (2010).
\bibitem{ouhd}S. Chen and J. Jing, Phys. Lett. B {\bf691}, 254 (2010); Phys. Rev. D {\bf82},084006 (2010).
\bibitem{ou15}M. Minamitsuji,  Phys. Rev. D {\bf 89}, 064025 (2014);
M. Minamitsuji,  Phys. Rev. D {\bf 89}, 064017 (2014).

\bibitem{ou17}T. Kolyvaris, G. Koutsoumbas, E. Papantonopoulos, G. Siopsis,  J. High Energy Phys. {\bf 11}, 133 (2013).

\bibitem{ou171}E. Babicheva and C. Charmousisa,  J. High Energy Phys. {\bf 08}, 106 (2014).
\bibitem{ou172}A. Anabalon, A. Cisterna and J. Oliva, Phys. Rev. D {\bf89}, 084050 (2014).


\bibitem{ou338}R. D. Daniels, G. M. Shore, Phys. Lett. B {\bf 367}, 75 (1996).
\bibitem{ou339} R. G. Cai, Nucl. Phys. B {\bf 524}, 639 (1998).
\bibitem{ou340}H. T. Cho, Phys.Rev. D {\bf 56},  6416 (1997).
\bibitem{ou341}V. A. De Lorenci, R. Klippert, M. Novello, J. M. Salim,  Phys. Lett. B {\bf 482}, 134 (2000).
\bibitem{ou3}S. Chen, J. Jing, J. Cosmo. Astro. Phys.{\bf10}, 002 (2015).

\bibitem{Ernst} F. J. Ernst, J. Math. Phys. {\bf17}, 54 (1976).
\bibitem{surf} W. J. Wild and R. M. Kerns, Phys. Rev. D {\bf 21}, 332 (1980).

\bibitem{hamin} B. Carter, Phys. Rev. {\bf174}, 1559 (1968).


\bibitem{dz42}X. Wu, T. Y. Huang, X. S. Wan, H. Zhang,  Astron. J. {\bf 133}, 1643 (2007).
\bibitem{dz43}D. Z. Ma, X. Wu, J. F. Zhu, New Astron. {\bf 13}, 216 (2008).
\bibitem{dz44}D. Z. Ma, X. Wu, and F. Y. Liu,  IJMPC, {\bf 19}, 1411 (2008).

\bibitem{Konoplya} R. A. Konoplya, Phys. Lett. B {\bf666}, 283 (2008).

\bibitem{lci2}G. Tancredi and A. S$\acute{a}$nchez,  Astron.J, {\bf 121}, 1171, (2001).
\bibitem{ch47} C. Froeschl$\acute{e}$, R. Gonczi and E. Lega,  Planetary $\&$ Space Science {\bf 47}, 881-886, (1997).

\bibitem{ch50} X. Wu, T. Y. Huang, H. Zhang,  Phys. Rev. D {\bf 74}, 083001 (2006).

\bibitem{an25} H. Poincar\'{e}, Rend. Circ. Mat. Palermo {\bf 33} 375 (1912).








\end{thebibliography}
\end{document}